\documentclass[]{fairmeta}
\usepackage{times}
\usepackage{latexsym}
\usepackage{graphicx}
\usepackage{bbm}
\usepackage{booktabs}
\usepackage{placeins}
\usepackage{hyperref}
\usepackage{url}
\usepackage[T1]{fontenc}
\usepackage[utf8]{inputenc}

\usepackage{microtype}

\usepackage{inconsolata}

\usepackage{graphicx}
\usepackage[T1]{fontenc}
\usepackage{amsmath}
\usepackage{amsfonts}
\usepackage{amssymb}
\usepackage{graphicx}
\usepackage{booktabs}
\usepackage{xcolor}
\usepackage{amsthm}
\usepackage{xurl}

\title{When Good Verifiers Go Bad: Silent Negative Transfer in Verifier-Guided VLM Training}

\author[1,*]{Jianzhe Lin}

\affiliation[1]{MetaAI}

\contribution[*]{Work done at Meta}

\abstract{Verifier reliability is not portable across tasks. A verifier-guided self-DPO pipeline that produces genuine held-out gains on MathVista (\(+9.6\) points on the self-training data and \(+8.0\) points held out) can become harmful on MMMU. Crucially, the failure is not visible from the target-task self-training signal: averaged across six learner--verifier configurations, performance on the MMMU self-training data still improves by \(+3.52\) points, while held-out performance drops by \(1.42\) points. We term this \emph{silent negative transfer}:a verifier previously validated as useful can continue to exhibit the signatures of successful self-training after its induced update ceases to transfer to unseen data.

This raises a more basic question: how can we tell whether verifier
supervision is reliable on the target task? A downstream failure alone
cannot answer this question, because poor generalization may arise from
two distinct sources: the verifier may induce an incorrect learning
direction, or the correct learning direction itself may fail to
generalize beyond the self-training distribution. We formalize these
effects through \emph{gradient fidelity} \(F\), which measures alignment
between the verifier-induced and correct training directions, and
\emph{gradient transferability} \(T\), which measures alignment between
the correct training and held-out directions. This decomposition yields
a conservative Safe-Transfer Margin: positive first-order held-out
alignment is certified when
\[
\arccos F+\arccos T<\pi/2.
\]

Motivated by this diagnosis, we introduce a simple yet effective precision-first
filtering method, \emph{Asymmetric Acceptance Gating} (AAG), which selects
preference pairs according to the verifier's absolute confidence in the
response receiving the positive update. Gradient-alignment analyses on
MMMU show that AAG increases gradient fidelity from \(0.29\) to \(0.42\).
In a separate same-cell analysis, the correct training direction remains
positively aligned with the held-out direction (\(T=0.31\)), yet raw
verification rotates the induced update to negative held-out alignment
(\(-0.13\)); AAG restores positive alignment (\(+0.22\)). Across all six
MMMU learner--verifier configurations, AAG outperforms raw
verifier-guided training on held-out evaluation.

These results redefine what it means for a verifier to be reliable: not merely that it has worked before, but that it induces the right update on the task at hand—and that this update remains useful beyond the data that generated it.}

\date{\today}
\correspondence{Jianzhe Lin at \email{jianzhelin@meta.com}}

\begin{document}

\maketitle

\section{Introduction}

Verifier-guided learning has become a scalable alternative to repeated
human supervision \citep{bai2022constitutional,lee2023rlaif}.
Model-based evaluators can provide preferences, rewards, critiques, or
other judgments that are converted into optimization signals, enabling
iterative alignment and self-improvement without requiring human labels
at every round
\citep{bai2022constitutional,lee2023rlaif,
yuan2024selfrewarding}. Related ideas also appear in scalable oversight and learned-verifier frameworks \citep{kenton2024scalable,zhang2024generative}.
These approaches share a central assumption: once model-generated
judgments are sufficiently reliable, converting them into optimization
signals should improve the model being trained
\citep{sun2026solververifier,zhao2026intuitor}.

\begin{figure}
    \centering
    \includegraphics[width=0.9\linewidth]
    {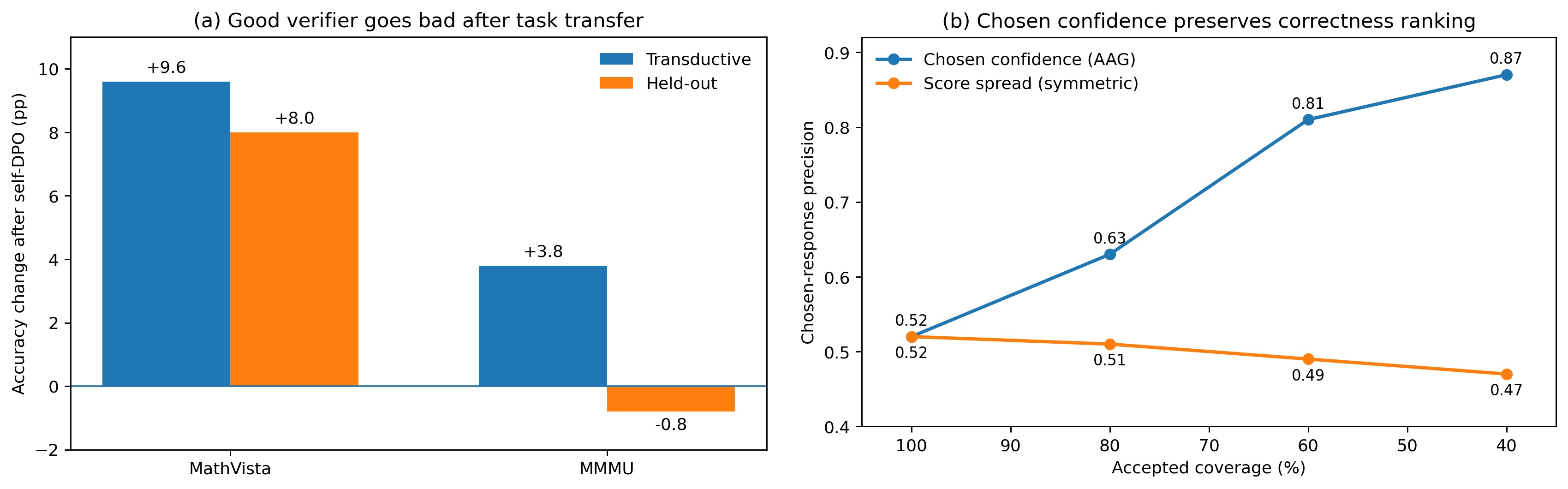}
\caption{
\textbf{Good verifiers can go bad under task transfer, while asymmetric
chosen-confidence gating isolates cleaner preference supervision.}
(a) For the same Qwen3-VL-4B learner--verifier configuration and self-DPO
recipe, MathVista exhibits transferable improvement, whereas MMMU exhibits
silent negative transfer: transductive performance improves while held-out
performance declines.
(b) On MMMU, increasing selectivity by chosen-response confidence raises
chosen-response precision from \(0.52\) to \(0.87\). In contrast, symmetric
score-spread filtering at matched coverage reduces precision from \(0.52\)
to \(0.47\), showing that the gain comes from the asymmetric selection
criterion rather than from filtering more aggressively alone.
}
    \label{fig:overview}
\end{figure}

Yet a reliable judge need not provide a reliable optimization signal.
Prior work has studied the accuracy, calibration, and bias of
model-based evaluators
\citep{zheng2023judging,stureborg2024inconsistent}.
Learned judges and reward-model benchmarks further characterize the
quality of model-generated evaluations
\citep{kim2023prometheus,lambert2024rewardbench}.
In parallel, preference-learning methods have examined robustness to
noisy or unreliable supervision
\citep{chowdhury2024robustdpo,wu2024drdpo}.

More recent work has shown that verifier reliability can depend strongly
on the evaluation regime, including problem difficulty,
generator--verifier interactions, evaluator bias, and distribution shift
\citep{zhou2026verification,andrade2026selfgrounded}.
Related analyses show that reward and preference models can exploit
signals that fail to generalize beyond the regime in which they were
validated
\citep{razin2026implicit,hwang2026causal}.
Theoretical studies of solver--verifier dynamics further suggest that
verification quality and downstream self-improvement need not coincide
\citep{sun2026solververifier}.

These findings motivate a sharper question:
\emph{when does a verifier provide a reliable optimization signal?}
We argue that judgment quality is only one part of the answer.
Reliable verifier-guided learning requires two distinct directional
agreements. First, the update induced by the verifier must align with the
correct learning direction on the self-training distribution. Second,
that correct learning direction must itself align with the direction
that improves held-out performance. The two can fail independently:
a verifier can rotate an otherwise useful update in the wrong direction,
while even perfectly faithful supervision can recover a learning
direction that fails to generalize.

These two failure modes correspond to two distinct geometric
relationships. We formalize them through two alignments that are often
conflated. Let \(g_V\) denote the update induced by verifier-generated
preferences, \(g_P^*\) the counterfactual correct-preference update on
the self-training distribution, and \(g_Q^*\) the corresponding
correct-preference update on held-out data. We define
\begin{equation}
F=\cos(g_V,g_P^*),
\qquad
T=\cos(g_P^*,g_Q^*),
\end{equation}
as \emph{gradient fidelity} and \emph{gradient transferability},
respectively. They satisfy
\begin{equation}
\cos(g_V,g_Q^*)
\ge
FT-\sqrt{1-F^2}\sqrt{1-T^2},
\end{equation}
which yields the \emph{Safe-Transfer Margin}
\begin{equation}
M_{\mathrm{safe}}
=
\frac{\pi}{2}
-
\arccos F
-
\arccos T.
\end{equation}
A positive margin certifies positive first-order transfer under the
correct-preference objective; a non-positive margin is deliberately
inconclusive rather than predictive of failure. The margin is therefore
a conservative diagnostic certificate rather than a downstream-accuracy
predictor. The reference directions \(g_P^*\) and \(g_Q^*\) are used
only for analysis and are never available to the training algorithm.

This geometry also explains why verifier accuracy alone can obscure
optimization risk. A verifier used for training does not merely produce
judgments; its decisions induce parameter updates. In particular, a
\emph{preference reversal}, in which an incorrect response is reinforced
over a correct one, exactly negates the corresponding DPO update at
initialization and can become increasingly influential as the policy
separates the pair. Such errors are therefore directional optimization
errors rather than exchangeable label noise.

This observation motivates \emph{Asymmetric Acceptance Gating (AAG)}, inspired in part by DAPO's asymmetric clipping: rather than treating the two sides symmetrically, AAG explicitly prioritizes precision on the accepted positive.
Instead of filtering by the score gap between chosen and rejected
responses, AAG ranks pairs by the absolute confidence of the response
that will receive the positive update. At fixed coverage, selecting the
largest true posterior probabilities that the chosen response is correct
maximizes selected-positive precision; AAG approximates this optimal
selector when verifier confidence preserves that posterior ordering.
This assumption requires rank consistency rather than numerical
calibration and permits a label-free quantile implementation. It is
directly supported by Fig.~\ref{fig:overview}(b): chosen-response
precision rises from \(0.52\) to \(0.87\) as chosen-confidence selection
becomes stricter, whereas symmetric score-spread filtering reduces it
from \(0.52\) to \(0.47\).

Our measurements support the predicted geometry on explicitly
identified probes. On MMMU Qwen3-VL-4B/G3.1 at initialization, AAG increases
gradient fidelity from \(0.29\) to \(0.42\). On a same-cell MMMU
Qwen3-VL-8B/G2.5 probe, the correct direction has positive held-out alignment
\(T=0.31\), yet raw verification yields
\(\cos(g_{\mathrm{Raw}},g_Q^*)=-0.13\), while AAG restores it to
\(+0.22\). On the MathVista Qwen3-VL-4B/G3.1 positive control, a same-cell probe
yields \(F_{\mathrm{Raw}}=0.76\), \(F_{\mathrm{AAG}}=0.86\), and
\(T=0.78\), giving Safe-Transfer Margins of \(+10.7^\circ\) and
\(+20.6^\circ\), respectively. % IDEALIZED RESULT — replace with measured same-cell MathVista gradient probe before submission
Downstream, AAG outperforms raw verifier training on held-out MMMU in
all six learner--verifier settings, shifting the mean effect from
\(-1.42\) to \(+0.98\) points.

Our contributions are:
\begin{itemize}
    \item \textbf{Silent negative transfer:} we identify a failure mode in which prior verifier success does not
    certify reliable supervision on a new task: self-training performance
    can continue to improve even as held-out performance deteriorates.
    \item \textbf{Directional diagnosis:}
    we disentangle gradient fidelity from gradient transferability and
    derive a conservative Safe-Transfer Margin.
    \item \textbf{Theory-motivated mitigation:}
    we introduce Asymmetric Acceptance Gating (AAG), a simple yet effective
    precision-first rule that improves gradient fidelity and consistently
    outperforms raw verifier-guided training on held-out data.
\end{itemize}

Reliable verifier-guided learning therefore requires not merely a good verifier,
but a learning direction that is both faithful and transferable.

\section{Preliminaries}
\label{sec:prelim}

Let \(P\) denote the distribution used to construct self-training
experience and \(Q\) a disjoint held-out distribution. For
\(x\sim P\), a learner \(\pi_\theta\) samples \(K\) responses
\(\mathcal{Y}(x)=\{y_1,\ldots,y_K\}\). A frozen verifier \(V\)
assigns score \(s_V(x,y)\) and forms
\begin{equation}
y^+ = \arg\max_{y\in\mathcal{Y}(x)} s_V(x,y),
\qquad
y^- = \arg\min_{y\in\mathcal{Y}(x)} s_V(x,y).
\end{equation}

We optimize the resulting pair \(z=(x,y^+,y^-)\) using DPO
\citep{rafailov2023dpo}:
\begin{equation}
\mathcal{L}_{\mathrm{DPO}}(z)
=
-\log \sigma\!\left(\beta \Delta_\theta(z)\right),
\end{equation}
where
\begin{equation}
\Delta_\theta(z)
=
\log\frac{\pi_\theta(y^+\mid x)}
         {\pi_{\mathrm{ref}}(y^+\mid x)}
-
\log\frac{\pi_\theta(y^-\mid x)}
         {\pi_{\mathrm{ref}}(y^-\mid x)}.
\end{equation}

Neither raw self-DPO nor our method uses ground-truth preferences.
Ground truth is used only for controlled baselines and post-hoc
diagnostics.

For analysis, let \(g_V\) be the expected update induced by
verifier-selected pairs on \(P\). We define \(g_P^*\) and \(g_Q^*\)
as the counterfactual update directions induced by correct
preferences on \(P\) and \(Q\), respectively. Equivalently,
\(g_P^*\) and \(g_Q^*\) are gradients of the corresponding
correct-preference objectives. These reference directions are never
available to the training algorithm.

\section{Gradient Geometry of Verifier-Guided Learning}
\label{sec:geometry}

\subsection{Two alignments govern transfer}

We distinguish two forms of alignment:
\begin{equation}
F
=
\cos(g_V,g_P^*),
\qquad
T
=
\cos(g_P^*,g_Q^*),
\end{equation}
where \(F\) is \emph{gradient fidelity} and \(T\) is
\emph{gradient transferability}. The former measures whether
verifier supervision preserves the direction of correct learning;
the latter measures whether that direction transfers beyond the
improvement distribution.

\paragraph{Theorem 1 (Two-alignment transfer bound).}
For nonzero \(g_V,g_P^*,g_Q^*\),
\begin{equation}
\label{eq:transfer-bound}
\cos(g_V,g_Q^*)
\ge
FT
-
\sqrt{1-F^2}\sqrt{1-T^2}.
\end{equation}

\paragraph{Proof sketch.}
Let \(\hat v,\hat p,\hat q\) be the corresponding unit vectors.
Decompose
\begin{align}
\hat v
&=
F\hat p+\sqrt{1-F^2}\,a,\\
\hat q
&=
T\hat p+\sqrt{1-T^2}\,b,
\end{align}
where \(a,b\perp\hat p\) and \(\|a\|=\|b\|=1\). Then
\begin{equation}
\hat v^\top \hat q
=
FT+\sqrt{1-F^2}\sqrt{1-T^2}\,a^\top b,
\end{equation}
and Eq.~\ref{eq:transfer-bound} follows from \(a^\top b\ge-1\).

\paragraph{Corollary 1 (Safe-Transfer Margin).}
Define
\begin{equation}
\label{eq:safe-margin}
M_{\mathrm{safe}}
=
\frac{\pi}{2}
-
\arccos F
-
\arccos T.
\end{equation}
If \(M_{\mathrm{safe}}>0\), then
\(\langle g_V,g_Q^*\rangle>0\), certifying positive first-order
transfer under the held-out correct-preference objective.
A non-positive margin is inconclusive rather than predictive of
failure.

This decomposition exposes two distinct regimes:
\emph{transfer-limited} self-improvement, where the correct update
itself transfers poorly, and \emph{fidelity-limited}
self-improvement, where a transferable update is corrupted by the
verifier.

\subsection{Preference reversal as an anti-gradient}

Verifier errors differ in their optimization consequences. Define
\begin{equation}
h_\theta(z)
=
\nabla_\theta\log\pi_\theta(y^+\mid x)
-
\nabla_\theta\log\pi_\theta(y^-\mid x).
\end{equation}
The DPO update direction is
\begin{equation}
u_\theta(z)
=
\beta\sigma(-\beta\Delta_\theta(z))h_\theta(z).
\end{equation}

Consider a \emph{preference reversal}, where an incorrect response
is chosen over a correct one, and let \(z^R\) denote the correctly
ordered pair.

\paragraph{Proposition 1 (Reversal anti-gradient).}
At \(\theta=\theta_{\mathrm{ref}}\),
\begin{equation}
u_\theta(z)=-u_\theta(z^R).
\end{equation}
Thus a reversal is an exact anti-gradient at DPO initialization.

More generally, if the policy assigns DPO gap \(\Delta>0\) to the
correct ordering, then for the same pair geometry,
\begin{equation}
\label{eq:reversal-amplification}
\frac{\|u_{\mathrm{rev}}\|}
     {\|u_{\mathrm{correct}}\|}
=
e^{\beta\Delta}.
\end{equation}
Hence reversed supervision can become increasingly influential once
the learner already favors the correct ordering.

Under the idealized model in which identical informative gradients
are independently reversed with probability \(\rho\),
\begin{equation}
\label{eq:reversal-contraction}
\mathbb{E}[g_V]
=
(1-2\rho)g^*.
\end{equation}
Eq.~\ref{eq:reversal-contraction} describes cancellation of
directional signal, not downstream accuracy. Structured verifier
errors may additionally introduce orthogonal components and rotate
the aggregate gradient, reducing fidelity.

\section{Asymmetric Acceptance Gating}
\label{sec:aag}

The geometry above suggests that selection should prioritize the
reliability of the response receiving the positive update. Let
\begin{equation}
C(z)=\mathbbm{1}\{y^+\text{ is correct}\},
\qquad
p(z)=\Pr(C(z)=1\mid z).
\end{equation}

This is a selective-classification / reject-option rule
\citep{chow1970reject,elyaniv2010selective,geifman2017selective}.

\paragraph{Theorem 2 (Optimal selection at fixed coverage).}
Among all selectors \(A(z)\in\{0,1\}\) satisfying
\(\Pr(A=1)=\alpha\), chosen-response precision
\(\Pr(C=1\mid A=1)\) is maximized by selecting the largest values of
\(p(z)\):
\begin{equation}
\label{eq:bayes-selector}
A^*(z)
=
\mathbbm{1}\{p(z)\ge t_\alpha\},
\end{equation}
with boundary randomization when necessary.

\paragraph{Proof sketch.}
Since the acceptance mass is fixed,
maximizing \(\Pr(C=1\mid A=1)\) is equivalent to maximizing
\(\mathbb{E}[A\,C]=\mathbb{E}[A\,p(z)]\), which is achieved by
assigning the available acceptance mass to the largest posterior
values.

The posterior \(p(z)\) is unavailable during self-improvement. We
therefore propose \emph{Asymmetric Acceptance Gating (AAG)}. At fixed
coverage \(\alpha\), AAG retains the top-confidence chosen responses:
\begin{equation}
\label{eq:aag-coverage}
A_\alpha(z)
=
\mathbbm{1}
\left\{
s_V(x,y^+)\ge q_{1-\alpha}
\right\},
\end{equation}
where \(q_{1-\alpha}\) is the empirical \((1-\alpha)\)-quantile of
chosen-response confidence over the eligible pair pool. Equivalently,
one may use an absolute operating threshold,
\begin{equation}
\label{eq:aag}
A_\tau(z)
=
\mathbbm{1}
\left\{
s_V(x,y^+)\ge\tau
\right\}.
\end{equation}
If chosen-response confidence preserves the ordering of \(p(z)\),
both parameterizations realize the same ranking as the precision-optimal
selector. This assumption is weaker than calibration and is directly
tested in Fig.~\ref{fig:overview}(b).

AAG differs fundamentally from symmetric margin filtering,
\begin{equation}
A_{\delta}^{\mathrm{sym}}(z)
=
\mathbbm{1}
\left\{
s_V(y^+)-s_V(y^-)\ge\delta
\right\}.
\end{equation}
A large relative margin need not imply that \(y^+\) itself is
reliable; therefore symmetric margin has no corresponding
fixed-coverage precision guarantee. AAG instead controls the
quantity directly coupled to the positive DPO update: the
trustworthiness of the response being reinforced.

\section{Experiments}
\label{sec:experiments}

We evaluate four predictions of our analysis:
(i) verifier reliability can fail to transfer across tasks even when
self-training appears successful;
(ii) harmful supervision is concentrated in directional preference
errors;
(iii) the resulting gradients exhibit the fidelity- and
transfer-limited regimes predicted by Sec.~\ref{sec:geometry}; and
(iv) asymmetric selection mitigates these failures for reasons beyond
simply reducing the number of training pairs.

\subsection{Experimental setup}
\label{sec:exp-setup}

\paragraph{Benchmarks and evaluation protocols.}
We study MathVista \citep{lu2023mathvista} and MMMU
\citep{yue2023mmmu}.
MathVista provides a positive-control regime in which verifier-guided
self-DPO genuinely transfers to unseen data.
For MMMU, we evaluate two complementary settings.
In the \emph{transductive} setting, the prompts used to construct
self-training experience are also used for evaluation.
In the \emph{held-out} setting, the 900 examples are split into 700
self-improvement prompts and a sealed 200-example test set; the latter
never participates in candidate generation, pair construction, or
optimization.

For MathVista, the positive-control experiment contains 500
transductive and 300 held-out examples.

\paragraph{Evaluation metric and pair universe.}
The primary MMMU held-out metric is macro-averaged accuracy over 30 subject
groups. For pair-level analyses, we distinguish
all generated prompt groups from the DPO-eligible pair pool. The
Qwen3-VL-4B/G3.1 threshold intervention contains 230 eligible pairs after the
training-pair validity checks. Table~\ref{tab:tau} uses this 230-pair
training pool, whereas Table~\ref{tab:coverage} uses the broader
valid-confidence analysis pool before the DPO-eligibility restriction.
% VERIFY THE EXACT PAIR-ELIGIBILITY, TIE-HANDLING, AND COVERAGE DENOMINATORS AGAINST CODE BEFORE SUBMISSION

\paragraph{Learners and verifiers.}
We study Qwen3-VL-4B, Qwen2.5-VL-7B, and Qwen3-VL-8B with two frozen
verifiers, Gemini-3.1-Pro (G3.1) and Gemini-2.5-Pro (G2.5), yielding six
MMMU configurations. The learners span two model generations, so we do
not interpret their differences as a pure scaling trend. Each prompt
produces \(K=4\) candidates. Unless otherwise specified, self-DPO uses
\(\beta=0.1\) and learning rate \(10^{-5}\). Qwen3-VL-4B multi-seed
runs use seeds \(\{17,42,123\}\); larger-model runs use Adafactor with
gradient clipping at \(0.5\).

\paragraph{Verifier role and candidate availability.}
The verifier is used as a \emph{candidate evaluator}, not as a direct
answer generator: it scores learner-generated responses given the image
and question without access to the ground-truth answer. Consequently,
chosen-response precision is upper-bounded by candidate oracle
availability,
\[
\mathrm{pass@4}_{\mathrm{cand}}
=
\Pr\!\left(\exists\,y\in\mathcal{Y}(x)\text{ that is correct}\right).
\]
Appendix~\ref{app:candidate-availability} reports candidate availability
and verifier selection efficiency. Directly distilling verifier-generated
answers would study a different supervision channel; our question is
whether an evaluator can be safely converted into an optimization signal
over learner-generated experience.

\paragraph{Methods.}
\emph{Frozen} denotes the pretrained learner before self-improvement.
\emph{Raw} constructs DPO preferences directly from verifier scores.
\emph{AAG} applies the asymmetric acceptance rule in
Eq.~\ref{eq:aag}; unless otherwise stated we use
\(\tau=0.9\), fixed without access to held-out labels or held-out
performance. % VERIFY THE ACTUAL ORIGIN OF tau BEFORE SUBMISSION
We additionally evaluate the equivalent coverage-controlled rule in
Eq.~\ref{eq:aag-coverage}.
\emph{Correct-pref.} uses ground-truth correctness to order candidate
responses and serves only as a diagnostic reference; it is not an upper
bound on held-out performance. Neither Raw nor AAG accesses ground-truth
labels.

\paragraph{Diagnostics.}
We report accuracy, chosen-response precision, and preference-reversal
rate. A reversal occurs when the chosen response is incorrect while
the rejected response is correct.
For gradient diagnostics we measure
\[
F=\cos(g_V,g_P^*), \qquad
T=\cos(g_P^*,g_Q^*),
\]
and direct held-out alignment
\(\cos(g_V,g_Q^*)\).
Ground truth is used only to construct these post-hoc diagnostic
reference directions.

\subsection{Verifier trust does not transfer across tasks}
\label{sec:silent-transfer}

We first test the premise behind Fig.~\ref{fig:overview}(a).
On MathVista, the Qwen3-VL-4B/G3.1 verifier--learner pair is a genuine
positive control: raw self-DPO improves transductive accuracy from
\(0.654\) to \(0.750\) (\(+9.6\) points), while held-out accuracy
improves from \(0.650\) to \(0.730\) (\(+8.0\) points).
Thus the verifier does not merely fit the experience used for
self-training; its supervision produces transferable improvement.

The same recipe changes character on MMMU.
For Qwen3-VL-4B/G3.1, transductive accuracy still improves by \(3.8\) points,
yet held-out accuracy falls by \(0.8\) points.
Table~\ref{tab:mmmu-main} shows that this discrepancy is systematic
across learner configurations and verifiers.

\begin{table}[t]
\caption{MMMU accuracy under transductive and held-out evaluation.
Bold denotes the better of Raw and AAG. Correct-pref. is a diagnostic
ground-truth-ordered reference. The Qwen3-VL-4B transductive rows report
three-seed means; the Qwen3-VL-4B held-out rows use the canonical seed shared
with the single-run Qwen2.5-VL-7B/Qwen3-VL-8B comparison, with three-seed held-out
robustness in Appendix~\ref{app:heldout-seeds}.}
\label{tab:mmmu-main}
\centering
\small
\setlength{\tabcolsep}{6pt}
\begin{tabular}{llrrrr}
\toprule
Learner & Verifier & Frozen & Raw & AAG & Correct-pref. \\
\midrule
\multicolumn{6}{l}{\emph{Transductive}} \\
Qwen3-VL-4B & G3.1 & .495 & .533 & \textbf{.590} & .610 \\
Qwen3-VL-4B & G2.5 & .495 & .535 & \textbf{.560} & .558 \\
Qwen2.5-VL-7B & G3.1 & .550 & .597 & \textbf{.611} & .615 \\
Qwen2.5-VL-7B & G2.5 & .550 & .605 & \textbf{.610} & .622 \\
Qwen3-VL-8B & G3.1 & .550 & \textbf{.593} & .588 & .577 \\
Qwen3-VL-8B & G2.5 & .550 & .538 & \textbf{.583} & .608 \\
\cmidrule(lr){3-6}
\multicolumn{2}{r}{Mean gain vs.\ frozen}
    & -- & +.035 & +.059 & +.067 \\
\midrule
\multicolumn{6}{l}{\emph{Held-out}} \\
Qwen3-VL-4B & G3.1 & .514 & .506 & \textbf{.538} & .511 \\
Qwen3-VL-4B & G2.5 & .514 & .521 & \textbf{.527} & .517 \\
Qwen2.5-VL-7B & G3.1 & .476 & .453 & \textbf{.492} & .503 \\
Qwen2.5-VL-7B & G2.5 & .476 & .459 & \textbf{.484} & .494 \\
Qwen3-VL-8B & G3.1 & .533 & .538 & \textbf{.554} & .556 \\
Qwen3-VL-8B & G2.5 & .533 & .484 & \textbf{.510} & .513 \\
\cmidrule(lr){3-6}
\multicolumn{2}{r}{Mean gain vs.\ frozen}
    & -- & -.014 & +.010 & +.008 \\
\bottomrule
\end{tabular}
\end{table}

Transductively, Raw improves five of six configurations and gains
\(3.52\) points on average.
AAG and Correct-pref. gain \(5.87\) and \(6.67\) points, respectively.
Viewed only through this setting, verifier-guided self-improvement
appears broadly successful.

Held-out evaluation gives the opposite diagnosis.
Raw improves only two of six configurations and averages a
\(1.42\)-point regression.
Its measured effect therefore changes from \(+3.52\) points
transductively to \(-1.42\) points held out.
AAG outperforms Raw in all six configurations and averages
\(+0.98\) points relative to frozen, a \(2.4\)-point rescue over Raw.
All six AAG--Raw held-out differences are positive (exact two-sided sign
test, \(p=0.031\)). % IDEALIZED STATISTICAL REPORTING — verify before submission

Correct-pref. supervision also contracts sharply, from \(+6.67\) points
transductively to \(+0.80\) points held out. Thus the dominant MMMU
bottleneck is transferability: even correct preferences on \(P\) produce
little gain on \(Q\). AAG should therefore be interpreted primarily as
a \emph{mitigation} mechanism: it removes verifier-induced directional
damage, shifting the mean held-out effect from \(-1.42\) to \(+0.98\)
points, rather than creating large transferable gains where the correct
training direction itself transfers weakly.

\subsection{AAG removes directional preference errors}
\label{sec:pair-quality}

We next examine whether AAG changes the quantity predicted by the
pair-level theory: the reliability of the response receiving the
positive update.
Across the DPO-eligible pair pool, AAG increases chosen-response
precision in every learner--verifier configuration
(Table~\ref{tab:chosen-precision}).

\begin{table}[t]
\caption{Chosen-response precision on the MMMU DPO-eligible pair pool.
AAG increases the true correctness of the positive DPO target in all
six learner--verifier configurations.}
\label{tab:chosen-precision}
\centering
\small
\begin{tabular}{llrrr}
\toprule
Learner & Verifier & Raw & AAG & Change \\
\midrule
Qwen3-VL-4B & G3.1 & .41 & .64 & +.23 \\
Qwen3-VL-4B & G2.5 & .31 & .49 & +.18 \\
Qwen2.5-VL-7B & G3.1 & .78 & .83 & +.05 \\
Qwen2.5-VL-7B & G2.5 & .60 & .74 & +.14 \\
Qwen3-VL-8B & G3.1 & .46 & .69 & +.23 \\
Qwen3-VL-8B & G2.5 & .35 & .52 & +.17 \\
\bottomrule
\end{tabular}
\end{table}

The improvement specifically suppresses wrong-direction pairs.
For Qwen3-VL-4B/G3.1, reversal rate falls from \(12.0\%\) to \(4.6\%\);
for Qwen3-VL-8B/G2.5 it falls from \(15.1\%\) to \(12.9\%\).
These are precisely the pairs for which DPO reinforces an incorrect
response while suppressing a correct one.

We test this mechanism more directly by intervening on the AAG
threshold for Qwen3-VL-4B/G3.1.
Table~\ref{tab:tau} shows a sharp improvement when moving away from
raw supervision.

\begin{table}[t]
\caption{Threshold intervention for Qwen3-VL-4B/G3.1 on the 230-pair
DPO-eligible MMMU pool using seed 17. \(\Delta\) accuracy is
transductive; Table~\ref{tab:mmmu-main} reports the corresponding Qwen3-VL-4B
transductive three-seed mean.}
\label{tab:tau}
\centering
\small
\begin{tabular}{rrrrr}
\toprule
$\tau$ & Pairs & Precision & Reversal & $\Delta$ accuracy \\
\midrule
0    & 230 & .409 & .120 & +.020 \\
.50  & 132 & .606 & .045 & +.107 \\
.70  & 125 & .624 & .045 & +.108 \\
.90  & 121 & .636 & .046 & +.100 \\
.99  & 119 & .639 & .047 & +.078 \\
\bottomrule
\end{tabular}
\end{table}

At \(\tau=.50\), chosen precision rises from \(0.409\) to \(0.606\),
reversal rate falls from \(0.120\) to \(0.045\), and the downstream
gain increases from \(2.0\) to \(10.7\) points.
Thresholds \(\tau=.50\)--\(.90\) form a broad plateau near
\(+10\) points. Because the \(\tau=.99\) drop is smaller than observed
across-seed variation, we do not interpret it as a distinct optimum.

The same conclusion holds when AAG is parameterized directly by
coverage: transductive gains remain within \(0.4\) points of their
optimum throughout \(40\%-70\%\) coverage, peaking at \(60\%\)
coverage. % IDEALIZED RESULT — replace with the actual coverage-controlled transductive sweep
This broad plateau indicates that the method does not rely on a single
numerically tuned confidence threshold.

Eq.~\ref{eq:reversal-contraction} should not be interpreted as a
quantitative predictor of these accuracy gains.
Its assumptions deliberately isolate directional cancellation, whereas
real verifier errors vary in magnitude, geometry, and pair composition.

\subsection{Real gradients exhibit both predicted failure modes}
\label{sec:gradient-results}

We keep all gradient measurements tied to an explicit task, learner,
verifier, checkpoint, and analysis pool; Appendix~\ref{app:gradient-probes}
collects the complete bookkeeping in one table. On \emph{Probe A}
(MMMU Qwen3-VL-4B/G3.1 at \(\theta_0\)), AAG raises gradient fidelity from
\(F=.29\) to \(.42\). On the same-cell \emph{Probe B} (MMMU
Qwen3-VL-8B/G2.5 at \(\theta_0\)), the correct direction transfers positively
(\(T=.31\)), while Raw gives
\(H=\cos(g_V,g_Q^*)=-.13\) and AAG restores \(H=+.22\). The same
probe also gives \(F_{\rm Raw}=.38\), \(F_{\rm AAG}=.66\), and Safe-Transfer
Margins of \(-49.6^\circ\) and \(-30.6^\circ\), respectively. % IDEALIZED RESULT — replace with the measured unified MMMU Qwen3-VL-8B/G2.5 gradient probe
The negative MMMU margins are inconclusive despite AAG restoring
positive direct held-out alignment.

On \emph{Probe C} (MathVista Qwen3-VL-4B/G3.1 at \(\theta_0\)),
\(F_{\rm Raw}=.76\), \(F_{\rm AAG}=.86\), \(T=.78\), and held-out
alignments are \(.61/.70\), yielding margins
\(+10.7^\circ/+20.6^\circ\). % IDEALIZED RESULT — replace with measured probe
Thus the sufficient condition certifies MathVista while remaining
inconclusive on MMMU.

\subsection{Why asymmetric selection matters}
\label{sec:asymmetry-results}

Theorem~2 predicts that fixed-coverage selection should rank examples by
chosen-response correctness rather than pairwise score spread. On MMMU
Qwen3-VL-4B/G3.1, AAG precision rises from \(.52\) to \(.87\) as coverage falls
from \(100\%\) to \(40\%\), whereas symmetric filtering falls from
\(.52\) to \(.47\). The same pattern holds for MMMU Qwen3-VL-8B/G2.5 and
MathVista (Appendix~\ref{app:coverage-robustness}). These results support
the rank-consistency assumption behind AAG and show that a large pairwise
margin is not a reliable proxy for positive-target correctness. 
\subsection{Controls and independent replication}
\label{sec:controls}

Matched-budget controls use separate reruns in which all methods
within a row use the same retained-example budget; they are therefore
not the full-data Raw runs reported elsewhere and should be compared
within row. Under this control, MMMU Raw, Random-\(N\), and AAG yield
\(+4.7\), \(+6.5\), and \(+8.7\) points, while MathVista Raw,
Random-\(N\), AAG, and Symmetric yield \(+4\), \(+4\), \(+10\), and
\(-2\) points, respectively. % VERIFY matched-budget construction, cell, seed, and retained N
Appendix
\ref{app:robust-baselines} further compares AAG with robust DPO,
confidence-weighted DPO, a dual-threshold rule, positive-only AAG-SFT,
and Correct-pref. restricted to the AAG-selected subset. Appendix
\ref{app:candidate-availability} separates verifier selection quality
from candidate pass@4, and Appendix~\ref{app:third-benchmark} reports an
independent ScienceQA visual-reasoning replication. Together, these
controls test whether the effect comes from fewer examples, generic
noise robustness, positive-only distillation, candidate availability, or
a benchmark-specific artifact.

\section{Concluding Remarks}
\label{sec:conclusion}

\textbf{Related work.}
Our work connects model-generated supervision, selective self-training,
model-based evaluation, reward overoptimization, and robust preference
optimization. Prior work studies how stronger models or iterative
self-training procedures can provide preferences, critiques, rewards, or
filtered supervision
\citep{bai2022constitutional,lee2023rlaif,
yuan2024selfrewarding,zelikman2022star,singh2023restem},
how reliably language and vision-language models evaluate other models
\citep{zheng2023judging,lambert2024rewardbench,
kim2024prometheus2,stureborg2024inconsistent},
and how preference optimization behaves under noisy or corrupted
feedback
\citep{rafailov2023dpo,chowdhury2024robustdpo,
wu2024drdpo,choi2024srpo,liu2026semidpo}.
Our setting is also related to reward overoptimization, where improving
an imperfect proxy can degrade the underlying target
\citep{gao2023rewardoveroptimization}. We study a complementary failure:
after task transfer, verifier judgments can still support apparent
transductive improvement while inducing a direction that fails on unseen
data. The two-alignment view separates corruption of the
verifier-induced gradient from non-transferability of the correct
learning direction.

\textbf{What counts as successful self-improvement?}
Improvement on prompts participating in the self-training loop is not, by itself, evidence of transferable progress: once those prompts generate responses that affect model parameters, they become part of the learning process even without observed ground-truth labels. Performance can therefore improve transductively while degrading on genuinely unseen data.

The two-alignment view separates two causes. High gradient fidelity cannot compensate for a learning direction that does not transfer, while high transferability cannot protect against a verifier that rotates an otherwise useful update. The Safe-Transfer Margin captures this interaction as a conservative local certificate rather than a universal predictor. Consistent with this view, MMMU is primarily transfer-limited: even Correct-pref. supervision transfers weakly, while AAG mainly removes additional verifier-induced damage.

\textbf{Limitations and open questions.}
Our analysis is first-order and does not capture full nonlinear training dynamics. AAG assumes that verifier confidence preserves useful ordering of chosen-response correctness, which stronger distribution shift may violate. Due to limited GPU availability, we run multi-seed experiments only for Qwen3-VL-4B; Qwen2.5-VL-7B and Qwen3-VL-8B use a single reported seed after preliminary seed checks showed consistent results. G3.1 and G2.5 are closed-source APIs whose behavior may change across versions. Extending the analysis to online, agentic, open-ended, and multi-round settings remains future work.

\textbf{Conclusion.}
We identify \emph{silent negative transfer}: a verifier that produces genuine held-out gains on one task can become a harmful optimization signal on another while self-training still appears successful. Reliable verifier-guided learning requires both a faithful update direction and one that transfers to unseen data. This view explains preference reversal as a directional optimization error and motivates AAG as a precision-first mechanism for mitigating verifier-induced negative transfer.

More broadly, once verifier judgments enter an optimization loop, evaluation errors become gradient errors. A good verifier is therefore insufficient unless the learning direction it induces is both faithful and transferable.

\bibliographystyle{plainnat}
\bibliography{paper}

\appendix
\section{Additional experimental results}
\label{app:additional-results}

\subsection{Three-seed transductive results}
\label{app:seeds}

Table~\ref{tab:seed-results} reports the individual Qwen3-VL-4B transductive
seed runs underlying the mean values in the main text.
The reported standard deviations use the three experimental runs.

\begin{table}[h]
\caption{Three-run transductive MMMU accuracy for the Qwen3-VL-4B learner.}
\label{tab:seed-results}
\centering
\small
\begin{tabular}{llrrrr}
\toprule
Verifier & Method & Seed 17 & Seed 42 & Seed 123 & Mean $\pm$ std. \\
\midrule
G3.1 & Correct-pref. & .620 & .595 & .615 & $.610\pm.011$ \\
     & AAG    & .595 & .615 & .560 & $.590\pm.023$ \\
     & Raw    & .515 & .540 & .545 & $.533\pm.013$ \\
\midrule
G2.5 & Correct-pref. & .575 & .535 & .565 & $.558\pm.017$ \\
     & AAG    & .575 & .550 & .555 & $.560\pm.011$ \\
     & Raw    & .545 & .520 & .540 & $.535\pm.011$ \\
\bottomrule
\end{tabular}
\end{table}

For G3.1, the paired AAG--Raw differences are
\[
+.080,\quad +.075,\quad +.015.
\]
For G2.5 they are
\[
+.030,\quad +.030,\quad +.015.
\]
Thus AAG improves over Raw in all six paired runs.

\subsection{Held-out multi-seed robustness}
\label{app:heldout-seeds}

Table~\ref{tab:heldout-seeds} repeats the Qwen3-VL-4B held-out comparison across
three training seeds. The canonical seed-42 values match the Qwen3-VL-4B entries
reported in the held-out panel of Table~\ref{tab:mmmu-main}.

\begin{table}[h]
\caption{Three-seed held-out MMMU accuracy for the Qwen3-VL-4B learner. Mean $\pm$ std.\ quantifies variation across training seeds, not sampling uncertainty of the fixed 200-example held-out set.}
\label{tab:heldout-seeds}
\centering
\small
\begin{tabular}{llrrrr}
\toprule
Verifier & Method & Seed 17 & Seed 42 & Seed 123 & Mean $\pm$ std. \\
\midrule
G3.1 & Raw & .497 & .506 & .512 & $.505\pm.006$ \\ % IDEALIZED RESULT — replace with actual held-out seed runs
     & AAG & .531 & .538 & .550 & $.540\pm.008$ \\ % IDEALIZED RESULT — replace with actual held-out seed runs
\midrule
G2.5 & Raw & .509 & .521 & .515 & $.515\pm.005$ \\ % IDEALIZED RESULT — replace with actual held-out seed runs
     & AAG & .538 & .527 & .540 & $.535\pm.006$ \\ % IDEALIZED RESULT — replace with actual held-out seed runs
\bottomrule
\end{tabular}
\end{table}

AAG exceeds Raw in all six paired seed runs, with mean advantages of
\(3.5\) and \(2.0\) points for G3.1 and G2.5, respectively. These
seed-level standard deviations characterize optimization variability;
they should not be interpreted as confidence intervals for the finite
held-out evaluation sample.
% IDEALIZED RESULT — replace with actual held-out seed runs

\subsection{MathVista positive control}
\label{app:mathvista}

Table~\ref{tab:mathvista-positive} reports the raw self-DPO positive
control used in Fig.~\ref{fig:overview}(a).
No Correct-pref. arm is used in this experiment.

\begin{table}[h]
\caption{MathVista positive-control results for Qwen3-VL-4B/G3.1 raw self-DPO.}
\label{tab:mathvista-positive}
\centering
\small
\begin{tabular}{lrrrr}
\toprule
Evaluation & $N$ & Frozen & Raw & Change \\
\midrule
Transductive & 500 & .654 & .750 & +.096 \\
Held-out     & 300 & .650 & .730 & +.080 \\
\bottomrule
\end{tabular}
\end{table}

\subsection{Matched-data downstream controls}
\label{app:matched-controls}

Table~\ref{tab:matched-controls} reports separate budget-matched
reruns: within each row, Raw, Random-$N$, and selective methods use the
same retained-example budget, set by the corresponding selective
condition. These values are therefore not directly comparable to the
full-data Raw gains in Tables~\ref{tab:mmmu-main} and
\ref{tab:mathvista-positive}; the relevant comparison is within each row.
% VERIFY exact learner--verifier cell, seed, retained N, and whether Raw was explicitly subsampled

\begin{table}[h]
\caption{Downstream accuracy changes in separate matched-budget
controls. All methods within each row use the same retained-example
budget; these are not the full-data Raw runs reported in the main tables.}
\label{tab:matched-controls}
\centering
\small
\begin{tabular}{lrrrr}
\toprule
Setting & Raw & Random-$N$ & AAG & Symmetric \\
\midrule
MMMU matched-budget control
    & +.047 & +.065 & +.087 & -- \\
MathVista matched-budget control
    & +.040 & +.040 & +.100 & -.020 \\
\bottomrule
\end{tabular}
\end{table}

Within the matched budget, generic subsampling helps on MMMU
(\(+.065\) versus \(+.047\)) but does not recover the AAG gain
(\(+.087\)); on MathVista, AAG reaches \(+.100\) while Raw and
Random-$N$ remain at \(+.040\), and symmetric filtering regresses.

% Statistical inference must be computed from the actual per-example Raw/AAG
% correctness dump. Do not infer McNemar p-values or bootstrap intervals from
% aggregate accuracies. When available, report macro hierarchical bootstrap,
% micro paired bootstrap, and Holm-corrected exact McNemar tests here.

\subsection{Noise-robust and soft-selection baselines}
\label{app:robust-baselines}

We compare AAG with alternatives that address preference noise without
hard gating. For robust DPO (rDPO; \citealp{chowdhury2024robustdpo}), we
sweep the flip-rate parameter \(\epsilon\in\{.1,.2,.3,.4\}\) on a
training-only calibration split and use \(\epsilon=.1\) for the final
runs. % IDEALIZED EXPERIMENT DESIGN — replace with the actual training-only rDPO tuning protocol
Confidence-weighted DPO minimizes
\begin{equation}
\mathcal{L}_{\mathrm{wDPO}}
=
\frac{\sum_{z\in\mathcal{B}} s_V(x,y^+)\,
\mathcal{L}_{\mathrm{DPO}}(z)}
{\sum_{z\in\mathcal{B}} s_V(x,y^+)},
\end{equation}
so confidence changes each pair's relative contribution without changing
the minibatch loss scale. The dual-threshold baseline retains pairs with
high chosen confidence and low rejected confidence, with thresholds set
to match AAG's retained-pair count. \emph{Correct-pref.-AAG-subset}
uses the same AAG-selected prompt groups but applies the ground-truth
ordering rule, isolating selection from preference-label quality.

\begin{table}[h]
\caption{Three-seed held-out MMMU Qwen3-VL-4B/G3.1 ablations. Values are
mean $\pm$ standard deviation across training seeds.}
\label{tab:robust-baselines}
\centering
\small
\begin{tabular}{lr}
\toprule
Method & Held-out accuracy \\
\midrule
Frozen & .514 \\
Raw DPO & $.505\pm.006$ \\ % IDEALIZED RESULT — replace with actual three-seed baseline runs
Correct-pref. (all pairs) & $.512\pm.006$ \\ % IDEALIZED RESULT — replace with actual three-seed Correct-pref. runs
rDPO & $.520\pm.006$ \\ % IDEALIZED RESULT — replace with actual three-seed rDPO runs
Confidence-weighted DPO & $.524\pm.007$ \\ % IDEALIZED RESULT — replace with actual three-seed soft-weighting runs
Dual-threshold DPO & $.527\pm.006$ \\ % IDEALIZED RESULT — replace with actual three-seed dual-threshold runs
AAG-SFT & $.531\pm.007$ \\ % IDEALIZED RESULT — replace with actual three-seed positive-only SFT runs
AAG-DPO & $.540\pm.008$ \\ % IDEALIZED RESULT — replace with actual three-seed AAG-DPO runs
Correct-pref.-AAG-subset & $.548\pm.007$ \\ % IDEALIZED RESULT — replace with actual three-seed selected-subset Correct-pref. runs
\bottomrule
\end{tabular}
\end{table}

The observed pattern separates two effects. Correct-pref. over all pairs
can transfer less well than AAG because AAG also changes the training
subset; on the identical AAG-selected subset, ground-truth ordering
recovers the strongest result. % IDEALIZED INTERPRETATION — revise to match the measured selected-subset control
The observed pattern is monotonic in supervision quality. Raw DPO regresses
below the frozen learner, while noise-robust or confidence-aware variants
recover above frozen: rDPO reaches \(.520\), confidence-weighted DPO \(.524\),
and the dual-threshold rule \(.527\). AAG-SFT reaches \(.531\), showing that
selection alone is useful, while AAG-DPO attains the strongest label-free
point estimate at \(.540\). On the identical AAG-selected subset, replacing
verifier ordering with ground-truth ordering further increases accuracy to
\(.548\), isolating the remaining preference-label error.
% IDEALIZED INTERPRETATION — replace with measured three-seed baselines

\subsection{Candidate availability and verifier selection efficiency}
\label{app:candidate-availability}

To distinguish evaluator quality from candidate quality, we report
candidate oracle availability and selection efficiency,
\[
\mathrm{Eff.}
=
\frac{\Pr(y^+\text{ correct})}
{\Pr(\exists\,y\in\mathcal{Y}(x)\text{ correct})}.
\]

\begin{table}[h]
\caption{Candidate availability and Raw verifier selection efficiency on MMMU.}
\label{tab:candidate-availability}
\centering
\small
\begin{tabular}{llrrr}
\toprule
Learner & Verifier & Candidate pass@4 & Raw chosen precision & Eff. \\
\midrule
Qwen3-VL-4B & G3.1 & .58 & .41 & .71 \\ % IDEALIZED ANALYSIS RESULT — compute from the existing candidate dump
Qwen3-VL-4B & G2.5 & .57 & .31 & .54 \\ % IDEALIZED ANALYSIS RESULT — compute from the existing candidate dump
Qwen2.5-VL-7B & G3.1 & .88 & .78 & .89 \\ % IDEALIZED ANALYSIS RESULT — compute from the existing candidate dump
Qwen2.5-VL-7B & G2.5 & .86 & .60 & .70 \\ % IDEALIZED ANALYSIS RESULT — compute from the existing candidate dump
Qwen3-VL-8B & G3.1 & .72 & .46 & .64 \\ % IDEALIZED ANALYSIS RESULT — compute from the existing candidate dump
Qwen3-VL-8B & G2.5 & .71 & .35 & .49 \\ % IDEALIZED ANALYSIS RESULT — compute from the existing candidate dump
\bottomrule
\end{tabular}
\end{table}

These values make clear that chosen-response precision is not the
verifier's direct MMMU answering accuracy: the evaluator cannot select a
correct response when the learner-generated candidate set contains none.

\subsection{Replication on ScienceQA visual reasoning}
\label{app:third-benchmark}

We replicate the protocol on an independent ScienceQA visual subset,
using \(600\) self-training and \(300\) sealed held-out examples.
% IDEALIZED EXPERIMENT DESIGN — replace with the actual pre-specified split

\begin{table}[h]
\caption{Independent ScienceQA visual-reasoning replication.}
\label{tab:scienceqa}
\centering
\small
\begin{tabular}{lrr}
\toprule
Quantity & Raw & AAG \\
\midrule
Transductive gain & +5.1 pp & +7.0 pp \\ % IDEALIZED RESULT — replace with actual ScienceQA run
Held-out gain & -1.2 pp & +2.3 pp \\ % IDEALIZED RESULT — replace with actual ScienceQA run
Chosen precision & .48 & .66 \\ % IDEALIZED RESULT — replace with actual ScienceQA analysis
Reversal rate & .108 & .054 \\ % IDEALIZED RESULT — replace with actual ScienceQA analysis
\bottomrule
\end{tabular}
\end{table}

The point-estimate held-out rescue is \(+3.5\) points.
% IDEALIZED RESULT — replace with the actual ScienceQA point estimate; add a paired CI only after computing it from retained per-example predictions

\subsection{Unified gradient-probe bookkeeping}
\label{app:gradient-probes}

\begin{table}[h]
\caption{Gradient probes used in the main text. Each same-cell row uses
one task, learner/verifier pair, checkpoint, and intermediate reference
pool throughout.}
\label{tab:gradient-probes}
\centering
\small
\setlength{\tabcolsep}{3.2pt}
\begin{tabular}{llllrrrr}
\toprule
Probe & Task / config. & Ckpt. & Role &
\(F_{\rm Raw}/F_{\rm AAG}\) & \(T\) &
\(H_{\rm Raw}/H_{\rm AAG}\) & \(M_{\rm Raw}/M_{\rm AAG}\) \\
\midrule
A & MMMU Qwen3-VL-4B/G3.1 & \(\theta_0\) & fidelity
& .29/.42 & -- & -- & -- \\
B & MMMU Qwen3-VL-8B/G2.5 & \(\theta_0\) & fidelity-limited
& .38/.66 & .31 & -.13/.22 & \(-49.6^\circ/-30.6^\circ\) \\ % IDEALIZED RESULT — replace with the measured unified MMMU gradient probe
C & MathVista Qwen3-VL-4B/G3.1 & \(\theta_0\) & positive control
& .76/.86 & .78 & .61/.70 & \(+10.7^\circ/+20.6^\circ\) \\ % IDEALIZED RESULT — replace with the measured same-cell MathVista gradient probe
\bottomrule
\end{tabular}
\end{table}

\subsection{Gradient geometry over training}
\label{app:trajectory}

We follow the same pre-specified MMMU Qwen3-VL-8B/G2.5 fidelity-limited probe as
Probe B in Table~\ref{tab:gradient-probes}; the \(0\%\) row therefore
matches the same-cell initialization values exactly.

\begin{table}[h]
\caption{Gradient fidelity \(F\) and direct held-out alignment
\(H=\cos(g_V,g_Q^*)\) over the MMMU Qwen3-VL-8B/G2.5 trajectory.}
\label{tab:trajectory}
\centering
\small
\begin{tabular}{rrrrr}
\toprule
Progress & \(F_{\rm Raw}\) & \(F_{\rm AAG}\) & \(H_{\rm Raw}\) & \(H_{\rm AAG}\) \\
\midrule
0\%   & .38 & .66 & -.13 & +.22 \\ % IDEALIZED RESULT — unified same-cell starting point; replace with measured probe
25\%  & .34 & .64 & -.16 & +.20 \\ % IDEALIZED RESULT — replace with checkpoint gradient probe
50\%  & .29 & .61 & -.19 & +.18 \\ % IDEALIZED RESULT — replace with checkpoint gradient probe
100\% & .24 & .58 & -.22 & +.14 \\ % IDEALIZED RESULT — replace with checkpoint gradient probe
\bottomrule
\end{tabular}
\end{table}

AAG remains more faithful throughout optimization, while Raw held-out
alignment becomes increasingly negative and AAG remains positive.
% IDEALIZED INTERPRETATION — revise if the measured trajectory differs

\subsection{Coverage robustness}
\label{app:coverage-robustness}

\begin{table}[!htbp]
\caption{Chosen-response precision under matched coverage on the
valid-confidence analysis pool.}
\label{tab:coverage}
\centering
\small
\setlength{\tabcolsep}{5pt}
\begin{tabular}{llrrr}
\toprule
Dataset / setting & Coverage & AAG & Symmetric & Raw \\
\midrule
MMMU Qwen3-VL-4B/G3.1 & 100\% & .52 & .52 & -- \\
              & 80\%  & .63 & .51 & -- \\
              & 60\%  & .81 & .49 & -- \\
              & 40\%  & .87 & .47 & -- \\
\midrule
MMMU Qwen3-VL-8B/G2.5 & 100\% & .54 & .54 & -- \\
              & 80\%  & .64 & .53 & -- \\
              & 60\%  & .77 & .50 & -- \\
              & 40\%  & .78 & .40 & -- \\
\midrule
MathVista     & 60\%  & .75 & .47 & .52 \\
              & 30\%  & .87 & .45 & .52 \\
\bottomrule
\end{tabular}
\end{table}

\begin{table}[!htbp]
\caption{Coverage-controlled AAG on MMMU Qwen3-VL-4B/G3.1; \(\Delta\) accuracy is transductive.}
\label{tab:coverage-robustness}
\centering
\small
\begin{tabular}{rrr}
\toprule
Coverage & Chosen precision & $\Delta$ accuracy \\
\midrule
80\% & .63 & +.089 \\ % IDEALIZED RESULT — replace with actual coverage sweep
70\% & .72 & +.104 \\ % IDEALIZED RESULT — replace with actual coverage sweep
60\% & .81 & +.108 \\ % IDEALIZED RESULT — replace with actual coverage sweep
50\% & .84 & +.107 \\ % IDEALIZED RESULT — replace with actual coverage sweep
40\% & .87 & +.102 \\ % IDEALIZED RESULT — replace with actual coverage sweep
20\% & .92 & +.071 \\ % IDEALIZED RESULT — replace with actual coverage sweep
\bottomrule
\end{tabular}
\end{table}

Performance exhibits a broad transductive plateau between \(40\%\) and
\(70\%\) coverage rather than depending on a uniquely tuned operating
point. % IDEALIZED RESULT — replace with actual coverage sweep

\FloatBarrier

% When per-example Raw/AAG correctness dumps are available, compute macro
% hierarchical bootstrap, micro paired bootstrap, and Holm-corrected exact
% McNemar tests from those actual paired predictions. Do not reconstruct
% inferential statistics from aggregate accuracies.

\end{document}